      \documentclass[prb,twocolumn,showpacs,
      preprintnumbers,amsmath,amssymb]{revtex4}

\usepackage{graphicx}
\usepackage{dcolumn}
\usepackage{bm}

\begin{document}


\title{Domain-wall structure of a classical Heisenberg ferromagnet on a M\"obius strip 
}

\author{Masanao Yoneya}
\author{Kazuhiro Kuboki}%
\email{kuboki@kobe-u.ac.jp}
\affiliation{%
Department of Physics, Kobe University, \\
Kobe 657-8501, Japan
}%
\author{Masahiko Hayashi}
\affiliation{%
Faculty of Education and Human Studies, 
Akita University, 
1-1 Tegatagakuen-machi, Akita, 010-8502, Japan\\
and JST-CREST, 4-1-8 Honcho, Kawaguchi, Saitama 332-0012, Japan
}%

\date{\today}

\begin{abstract}
We study theoretically the structure of domain walls in ferromagnetic states 
on M\"obius strips. A two-dimensional classical Heisenberg ferromagnet 
with single-site anisotropy is treated within a mean-field approximation by 
taking into account the boundary condition to realize the M\"obius geometry. 
It is found that two types of domain walls can be formed, namely, parallel or 
perpendicular to the circumference, and that the relative stability of these 
domain walls is sensitive to the change in temperature and 
an applied magnetic field. The magnetization has a discontinuity as a function 
of temperature and the external field. 
\end{abstract}

\pacs{75.60.Ch, 75.10.Hk, 75.60.Ej
}
\maketitle

\section{Introduction}

The effect of system geometry on the physical properties has 
stimulated  renowned interest after the realization of crystals with 
unusual shapes, {\it e.g.}, ring, cylinder 
and especially M\"obius strip,\cite{Tanda1,Tanda2,Okajima,Tanda} because  
the ordered states in these systems could be quite different from those 
in the ordinary bulk systems. These single crystals are made of 
quasi-one-dimensional conductors, such as, NbSe$_3$. 

Hayashi and Ebisawa\cite{Hayashi-Ebisawa} 
studied the $s$-wave superconducting (SC) states
on a M\"obius strip based on the Ginzburg-Landau (GL) theory, and found that 
the Little-Parks oscillation is modified compared to that for the ordinary ring-shaped 
samples. When the flux quanta threading the M\"obius strip 
are close to half-odd integer times flux quantum $\phi_0 =hc/2e$ 
($h$, $c$, and $-e$ being the Planck constant, the speed of light and the electron 
charge, respectively), a state appears in which the real-space node in SC gap 
exists along the circumference of the strip. Later a more detailed examination using 
the Bogoliubov-de Gennes (BdG) theory confirmed this prediction.\cite{HEK} 
Charge density wave (CDW) states in ring-shaped crystals were also examined by 
Nogawa and Nemoto, \cite{Nogawa} and Hayashi {\it et al}.\cite{HEKdis,HEKcdw}
There the effect of frustration due to bending of the lattice 
was clarified. 
The persistent current in a M\"obius strip as a function of the applied flux was 
examined by Yakubo {\it et al}.\cite{Yakubo} and Mila {\it et al}.\cite{Mila}
Wakabayashi and Harigaya\cite{Wakabayashi-Harigaya} 
studied the M\"obius strip made of a nanographite 
ribbon, and the effect of M\"obius geometry on the edge localized states has 
been explored. 
Kaneda and Okabe \cite{Kaneda-Okabe} studied the 
Ising model on M\"obius, especially the effect of sample geometry on the finite scaling properties.

In this paper we study the ferromagnetic states on M\"obius strips. 
Although the single crystals with M\"obius geometry have been 
obtained for systems with CDW or superconducting order 
up to now, we expect that the synthesis of ferromagnetic M\"obius strips 
using quasi-one-dimensional ferromagnets could be possible. 
In such a system there should be a domain wall (DW), and two kinds of DWs, 
{\it i.e.}, those parallel and perpendicular to the circumference are expected. 
We will show that these two types of DWs can actually exist. 
An interesting point is that the relative stability of these 
two kinds of DWs is quite sensitive to the change in temperature and 
the applied magnetic field. We expect that this might be used in the field 
of technological applications in future. 

This paper is organized as follows. 
In Sec. II the model and the method of calculations are presented. 
We study the structure of domain walls on M\"obius strips in Sec. III. 
In Sec. IV the effect of an applied magnetic field is examined. 
Section V is devoted to summary.

\section{Model and Mean-field approximation}

In order to study the ferromagnetic states on M\"obius strips 
we consider a classical Heisenberg model 
with single-ion anisotropy, since this model is most suitable to examine the spatial 
variations of the magnetization.  
\begin{figure}[htb]
\begin{center}
\includegraphics[width=7cm,clip]{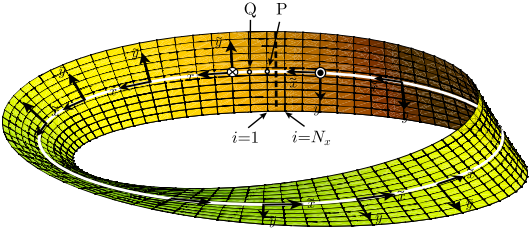}
\caption{Schematic picture of a M\"obius strip.  
The axes are those for the relative coordinate system at each point. 
See text for details. 
}
\label{dos}
\end{center}
\end{figure}
A model system with M\"obius geometry is constructed as follows.  
We consider $N_y$ (even integer) sets of ferromagnetic chains 
 (each chain has $N_x$ sites; $N_x$ being an integer), 
and these chains are weakly coupled ferromagnetically.  
We denote the $i$ th spin on the $j$ th chain ($1 \leq i \leq N_x$, 
$1 \leq j \leq N_y$)  as ${\bf S}_{i,j}$ ($|{\bf S}_{i,j}| = 1$). 
As a first step we form a cylinder, where the circumference is parallel to the chains. 
If one moves one site along the chain, the direction normal to the strip 
changes by the angle $\theta_1 \equiv 2\pi/N_x$. 
Then we twist the strip along the circumference at the middle of the strip 
(shown as a white curve in Fig.1) so that the twist angles at the neighboring 
sites on the white curve (for example, the points denoted as $P$ and $Q$) differ by 
$\theta_2 \equiv \pi/N_x$.
Finally the boundary condition is imposed,  
\begin{equation} 
\displaystyle {\bf S}_{N_x+1,j} = {\bf S}_{1,N_y+1-j},  
\end{equation}
to get a M\"obius strip. 
The direction normal to the strip is different from site to site, 
so we define the absolute coordinate system $(x,y,z)$ and 
the relative coordinate system $({\tilde x}, {\tilde y}, {\tilde z})$ at each point  
and use one of them which is better to describe the particular situation.  
For the former, the point with $i=1$ on the middle of the strip ($P$ in Fig.1)
is taken to be the origin, 
and the $z$ direction is perpendicular to the strip. The $x$ and $y$ directions are 
within the strip, and the former (latter) is  parallel (perpendicular) to the circumference. 
For the relative coordinate system, 
the ${\tilde z}$ direction is defined as that perpendicular to 
the strip at each point, and ${\tilde x}$ (${\tilde y}$) direction is 
parallel (perpendicular) to the circumference at that point (see Fig.1). 
Namely, the relative coordinate system for a particular point P is defined as 
the absolute coordinate system.
Then the spin in the relative coordinate system at a site $(i,j)$, 
${\bf {\tilde S}}_{i,j}$, is expressed by that in the absolute coordinate 
system, ${\bf S}_{i,j}$,  as  
\begin{equation}\begin{array}{rl}
{\tilde S}^x_{i,j} = & \displaystyle 
S^x_{i,j}\cos(k\theta_1) - S^z_{i,j}\sin(k\theta_1),  \\
{\tilde S}^y_{i,j} = & \displaystyle 
S^y_{i.j}\cos(k\theta_2) \\
& -  \displaystyle (S^x_{i,j}\sin(k\theta_1)+S^z_{i,j}\cos(k\theta_1))\sin(k\theta_2),  \\
{\tilde S}^z_{i,j} = & \displaystyle
(S^z_{i,j} \cos (k\theta_1) + S^x_{i,j} \sin(k\theta_1)) \cos(k\theta_2)  \\
& + \displaystyle S^y_{i,j} \sin (k\theta_2),   
\end{array}\end{equation} 
with $ k=i-1$.
Now the Hamiltonian of our system is given as 
\begin{equation}\begin{array}{rl}
 {\cal H} = & \displaystyle -J \sum_{i,j} 
({\bf S}_{i,j}\cdot{\bf S}_{i+1.j}
+ \alpha {\bf S}_{i,j}\cdot{\bf S}_{i,j+1}) \\
& \displaystyle -D\sum_{i,j} ({\tilde S}_{i,j}^z)^2 
- {\bf H}\cdot \sum_{i,j} {\bf S}_{i,j}, 
\end{array}\end{equation}
where $J (>0)$ and $J\alpha$ are the exchange energies between 
nearest-neighbor spins located along and perpendicular to the chains, 
respectively ($0 < \alpha < 1$). 
$D$ is the single-ion anisotropy energy which favors spin alignment 
perpendicular to the strip if $D > 0$ and ${\bf H}$ is the external magnetic field. 

We treat ${\cal H}$ using a mean-field (MF) approximation and the 
resulting MF Hamiltonian reads  
\begin{equation}\begin{array}{rl}
{\cal H}_{MF} = & \displaystyle \sum_{i,j} {\cal H}_{i,j} + E_0, \\
{\cal H}_{i,j} = & \displaystyle - ({\bf m}_{i,j} + {\bf H}) \cdot 
{\bf S}_{i,j} - D \big({\tilde S}_{i.j}^z\big)^2 , 
\end{array}\end{equation}
where ${\bf m}_{i,j}$ and $E_0$ are the mean-field acting on the spin ${\bf S}_{i,j}$  
and the constant term of the energy, respectively, 
\begin{equation}\begin{array}{rl}
 {\bf m}_{i,j} = & \displaystyle¡¡
J \sum_{\delta=\pm 1} (\langle {\bf S}_{i+\delta,j}\rangle 
+ \alpha \langle {\bf S}_{i,j+\delta}\rangle),  \\
 E_0 = & \displaystyle J \sum_{i,j} \langle {\bf S}_{i,j} \rangle \cdot  
\sum_{\delta=\pm 1} (\langle {\bf S}_{i+\delta,j} \rangle 
+ \alpha \langle {\bf S}_{i,j+\delta} \rangle),  
\end{array}\end{equation}
with the boundary condition 
$\displaystyle \langle {\bf S}_{i,0} \rangle = \langle {\bf S}_{i,N_y+1} \rangle  = 0$. 
The spins ${\bf S}_{i,j}$ are expressed by the angles $\theta_{ij}$ and $\phi_{ij}$; 
$\displaystyle S^x_{i,j} = \sin \theta_{ij} \cos \phi_{ij}$, 
$\displaystyle S^y_{i,j} = \sin \theta_{ij} \sin \phi_{ij}$, 
$\displaystyle S^z_{i,j} = \cos \theta_{ij}$, 
\begin{equation}\begin{array}{rl}
\langle {\bf S}_{i,j} \rangle 
= & \displaystyle \frac{1}{Z_{i,j}} {\rm Tr}_{i,j} 
{\bf S_{i,j}} e^{-\beta {\cal H}_{i,j}} , \\ 
& \\
 Z_{i,j} =  & \displaystyle {\rm Tr}_{i,j} e^{-\beta {\cal H}_{i,j}}, \\
& \\
\displaystyle {\rm Tr}_{i,j} (\cdots) = & \displaystyle
\frac{1}{4\pi} \int_0^{2\pi} d\phi_{ij} \int_0^\pi d\theta_{ij} 
\sin \theta_{ij} (\cdots) 
\end{array}\end{equation} 
with $\beta = 1/T$. ($T$ is the temperature and we use units $k_B = 1$.)
The free energy is  calculated as 
\begin{equation} 
\displaystyle F=  E_0 - T \sum_{i,j}\log Z_{i,j}. 
\end{equation}
Equations (4)-(6) constitute the self-consistency equations which we solve  
numerically by the method of iteration. 
If we obtain several solutions to satisfy the self-consistency equations 
for the same set of parameters in the model, the solution with the 
lowest free energy is the true one. 

Because of the anisotropy in spin space due to the presence of the $D$ term, 
Mermin-Wagner theorem\cite{MW} does not apply to our system.
Then the finite $T_C$ would be expected and our mean-field treatment 
strictly in two-dimensional systems should be justified in a qualitative sense. 
In order to relate our model to real samples which have finite width 
perpendicular to ${\tilde x}$ and ${\tilde y}$ directions, 
we assume that there is a weak three dimensionality 
(a finite but thin width)   
and that the system is uniform along the third direction. 
These assumptions allow us to 
consider the realistic systems by use of the model presented above. 

\bigskip

\section{structure of domain wall}

In this section we examine the spatial variations of the magnetization  
$\langle S^\mu_{i,j} \rangle$ ($\mu = x,y,z$) in 
order to clarify the structure of DWs on M\"obius strips. 
In the following we fix the system size to be $N_x = 100$ and $N_y = 10$, 
and take $J=1$ as the unite of energy. 
First we show the $i$ and $j$ dependences (direction parallel and perpendicular 
to the circumference, respectively) for $D=0.5$, $\alpha=0.04$ and 
$T=0.2T_C$  ($T_C=0.811J$) in Fig.2. 
As seen in Figs.2(a) and 2(b)  $\langle S^\mu_{i,j} \rangle$ [$\mu = x,y,z$; 
note that $(x,y,z)$ is the absolute coordinate system]  
for $i = N_x$ in the $j=1$ chain connect smoothly 
to those for $i=1$ in the $j=10$ chain. 
On the contrary in Fig.2(c) there is a 
sign change in $\langle S^y_{5,j} \rangle$ in the middle of the strip 
(between $j=5$ and $j=6$) and this kind of sign change occurs for all $i$. 
\begin{figure}[htb]
\begin{center}
\includegraphics[width=6cm,clip]{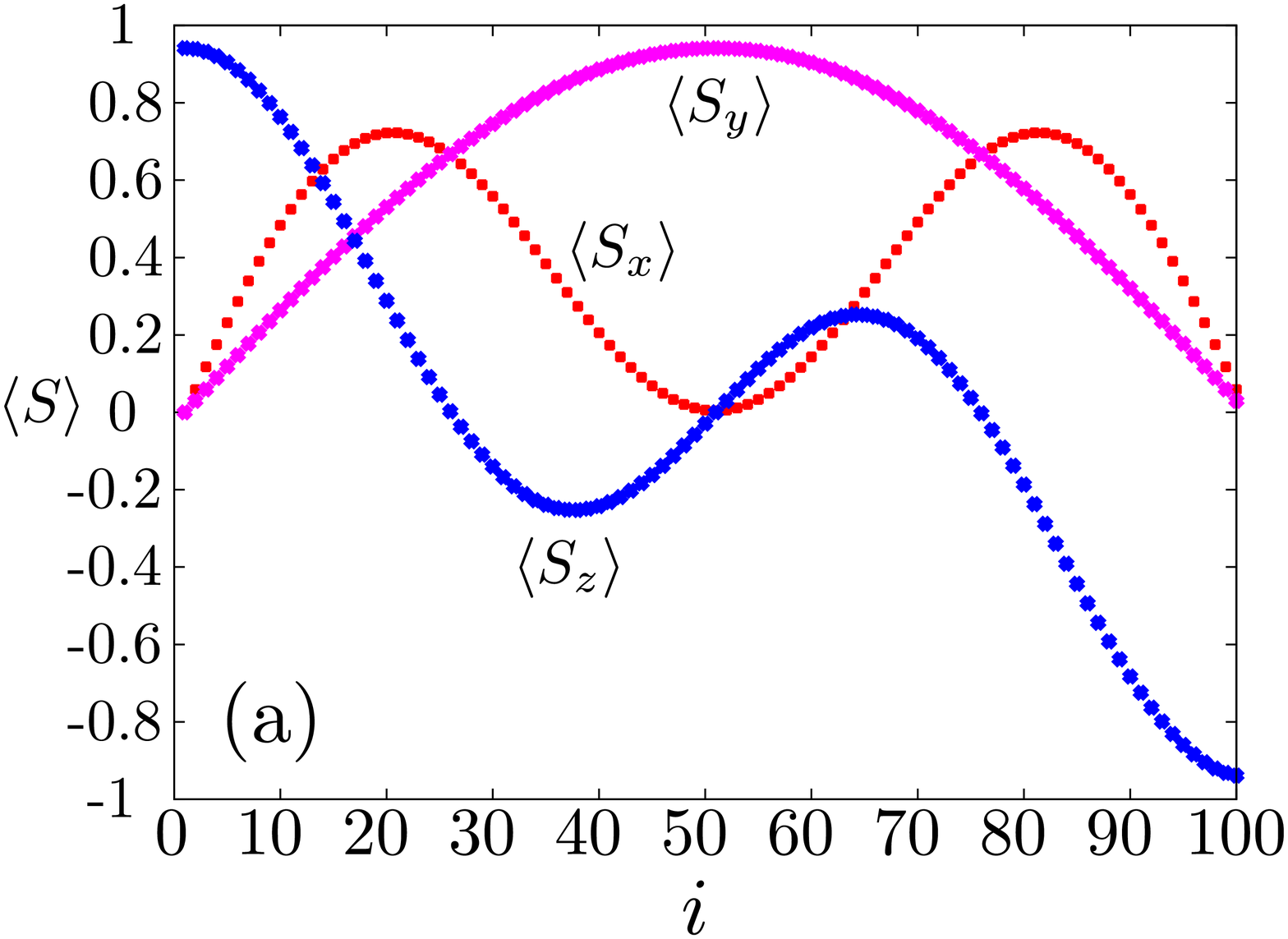}
\includegraphics[width=6cm,clip]{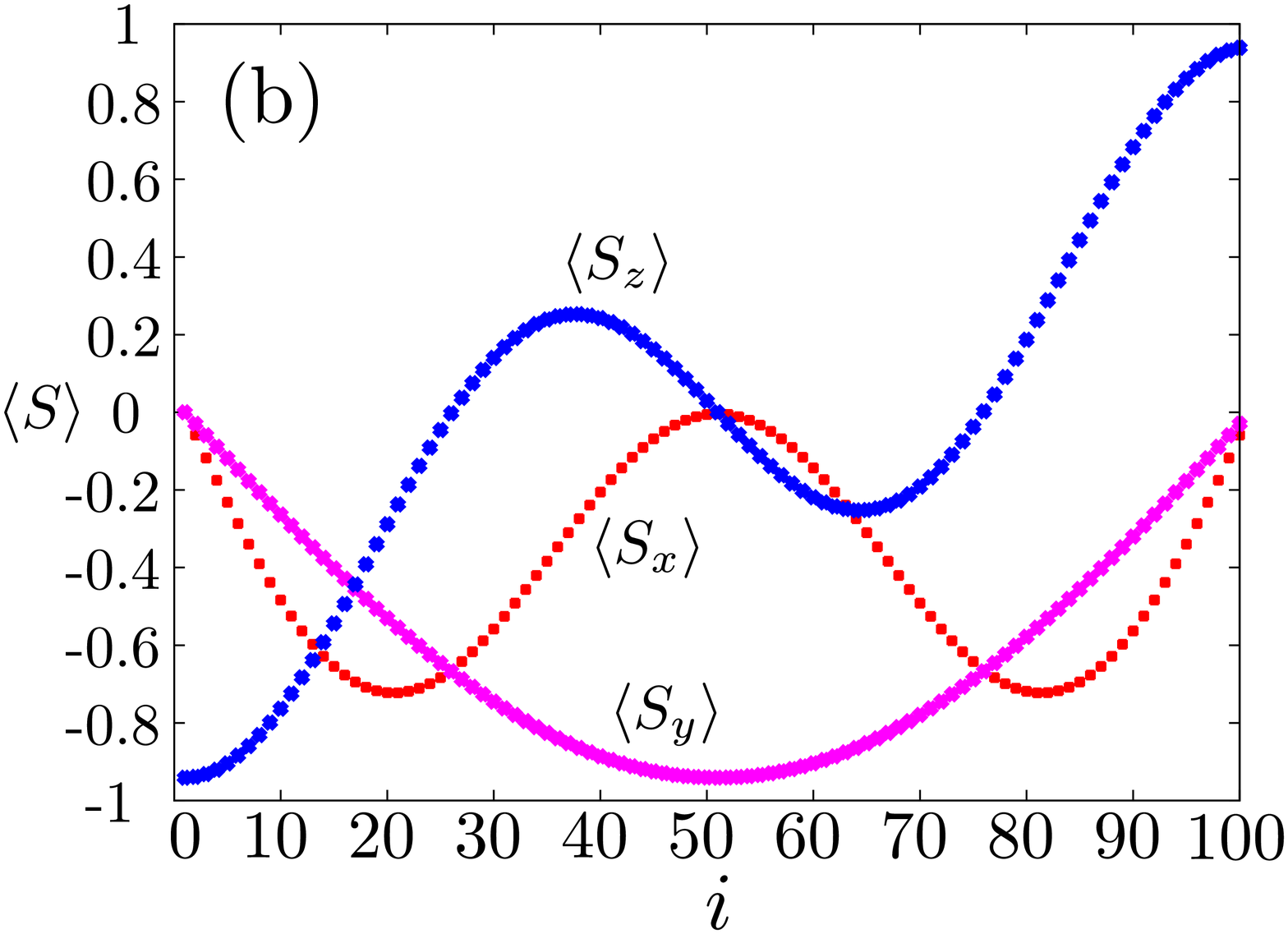}
\includegraphics[width=6cm,clip]{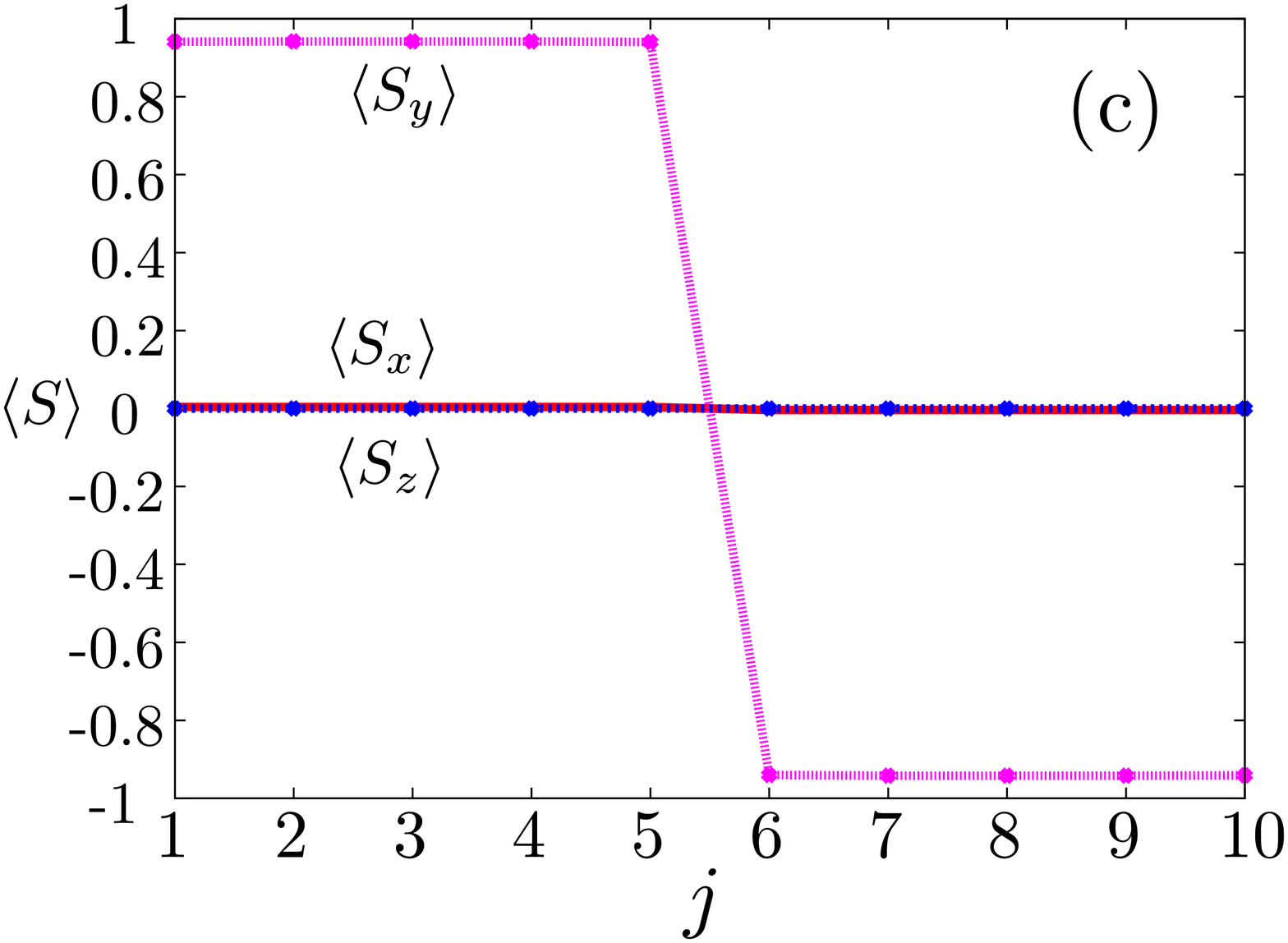}
\caption{
The spatial variations of $\langle {\bf S}_{i,j}\rangle$ for 
$D=0.5$, $\alpha=0.04$ and $T=0.2T_C$. 
 (a) $j=1$, (b) $j=10$
and (c) $i=51$. Note that $(x,y,z)$ is the absolute coordinate system.}
\end{center}
\end{figure}
It means that a domain wall is formed along the 
circumference of the M\"obius strip as schematically shown in Fig.3(a), 
and we call this kind of DW as type-A . 
When we transform $S^\mu_{i,j}$ (absolute coordinate)
to ${\tilde S}^\mu_{i,j}$ (relative coordinate), the latter has only ${\tilde z}$ 
component except near the DW.  
\begin{figure}[htb]
\begin{center}
\includegraphics[width=8cm,clip]{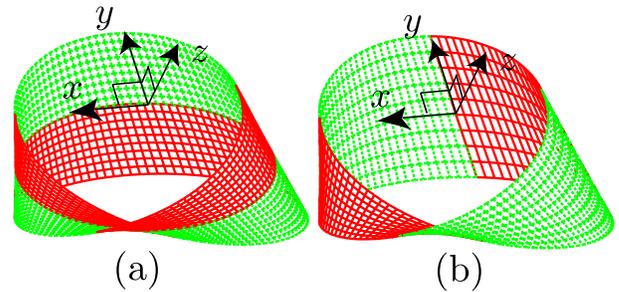}
\caption{Two types of domain walls: (a) type A - parallel to the circumference and 
(b) type B - perpendicular to it. $(x,y,z)$ are the axes of the absolute coordinate system.}
\label{dos}
\end{center}
\end{figure}

Next we consider the case of stronger inter-chain coupling, $\alpha = 0.15$. 
In Fig.4(a) the $i$ dependence of $\langle S^\mu_{i,j} \rangle$ is  
shown for $j=1$ and $j=10$ for $T=0.2T_C$ ($T_C=0.883J$). 
(The results for $j=1$ and $j=10$ are the same.)
In this case  $\langle S^\mu_{i,j} \rangle$ 
for $i = N_x$ in the $j=1$ chain do not connect smoothly to those for 
$i=1$ in the $j=10$, and so the discontinuity 
is observed. On the other hand the $j$ dependence is smooth [Fig.4(b)],  
and this behavior is the same for all $i$.
\begin{figure}[htb]
\begin{center}
\includegraphics[width=6cm,clip]{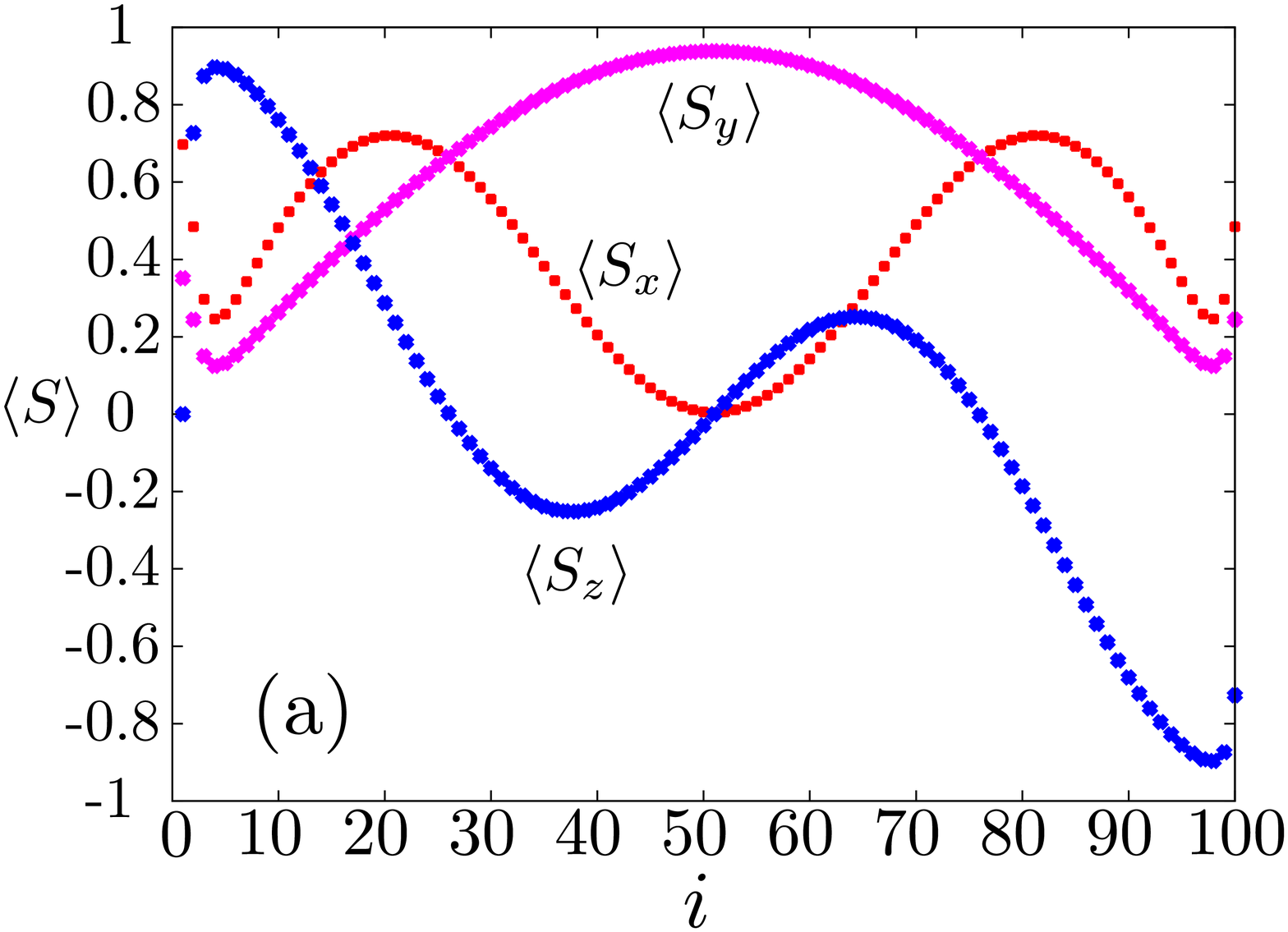}
\includegraphics[width=6cm,clip]{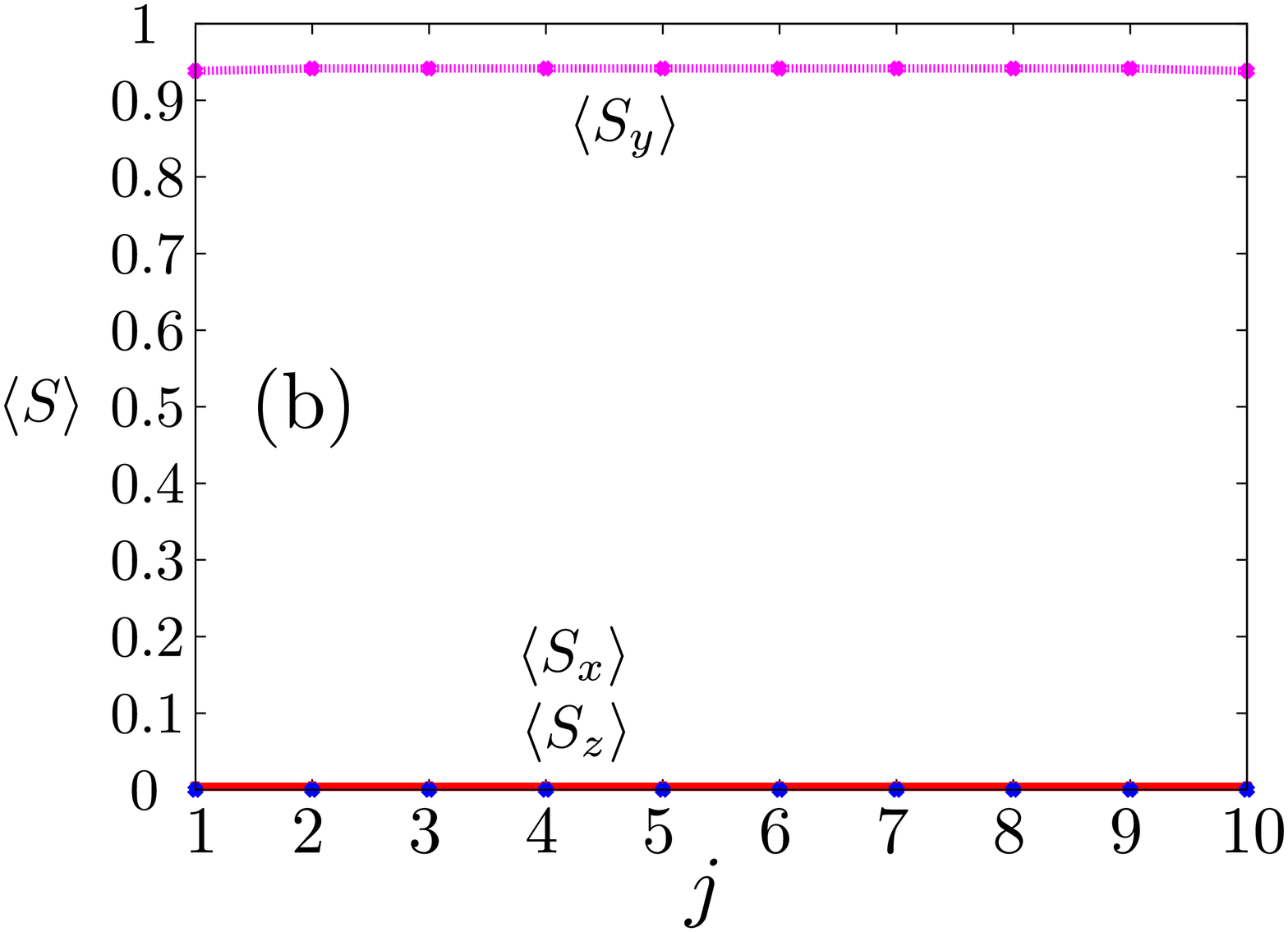}
\caption{
The spatial variations of $\langle {\bf S}_{i,j}\rangle$ for 
$D=0.5$, $\alpha=0.15$ and $T=0.2T_C$.  (a) $j=1$, 
and (b) $i=51$.}
\end{center}
\end{figure}
It indicates that the DW is located 
perpendicularly to the circumference in contrast to the case of $\alpha = 0.04$, 
and we call this kind of DW as type B [schematically shown in Fig.3(b)]. 
Naively we expect that a type-A (B) domain wall 
is stable if  $N_x \alpha < N_y$ ($N_x \alpha > N_y$),  
since the energy to create a DW is smaller compared to that for 
type B (A). The results shown in Figs.2 and 4 are consistent with this expectation. 

However, this stability condition for DWs will not hold at higher temperatures
or reduced values of $D$. 
In Fig.5 the phase diagram in the plane of $\alpha$ and $T$ is shown.  
\begin{figure}[htb]
\begin{center}
\includegraphics[width=6cm,clip]{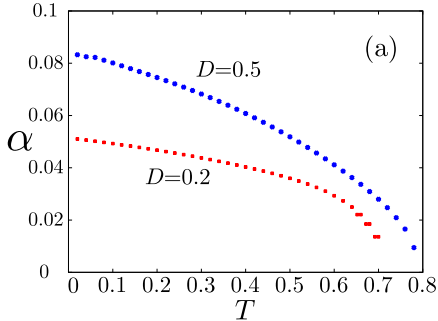}
\caption{
Phase diagram in the plane of $T$ and $\alpha$. 
Curves are the boundaries between type-A and B domain walls, 
and the region close to the origin corresponds to the type-A domain wall. }
\end{center}
\end{figure}
\begin{figure}[htb]
\begin{center}
\includegraphics[width=6cm,clip]{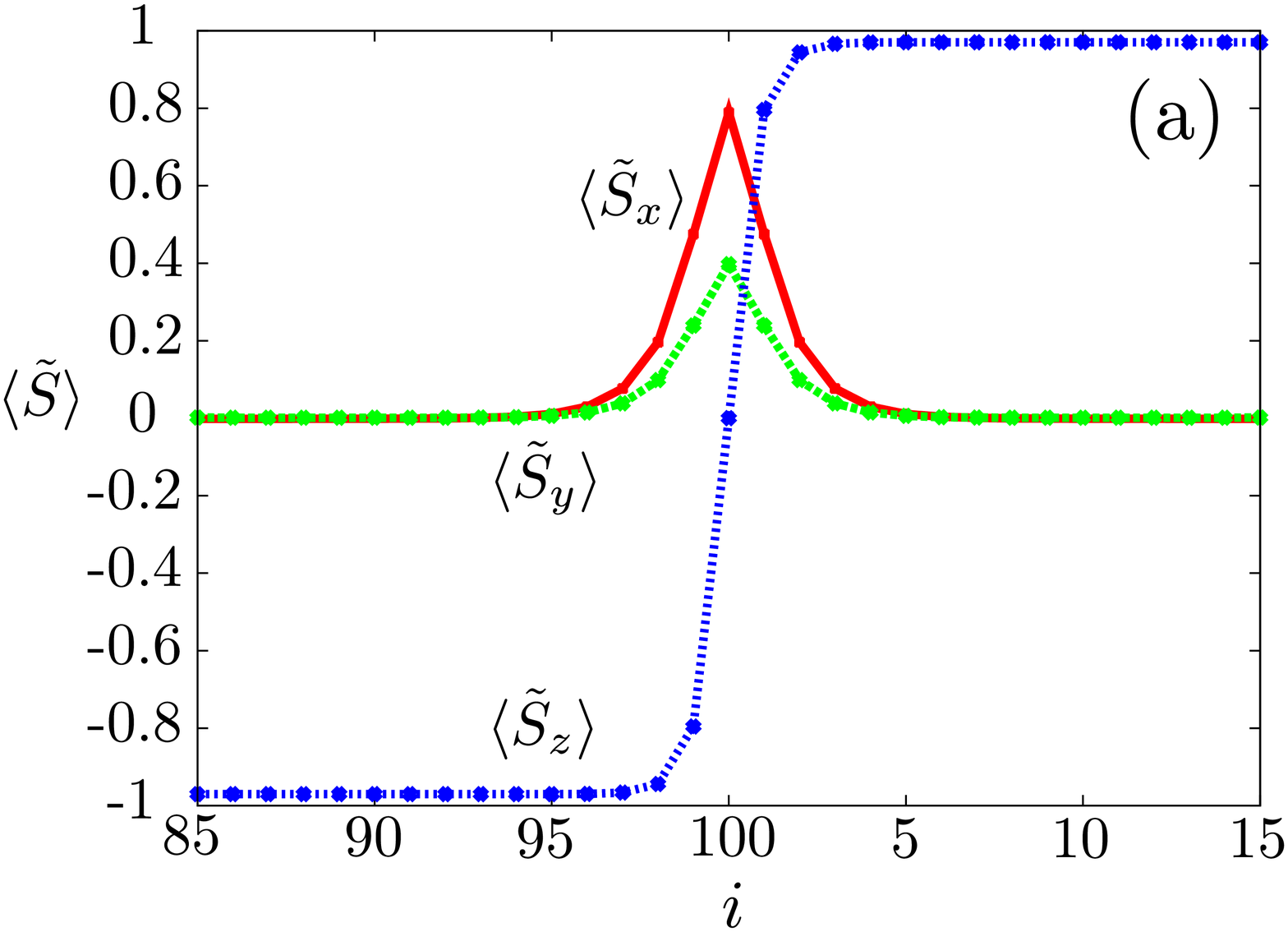}
\includegraphics[width=6cm,clip]{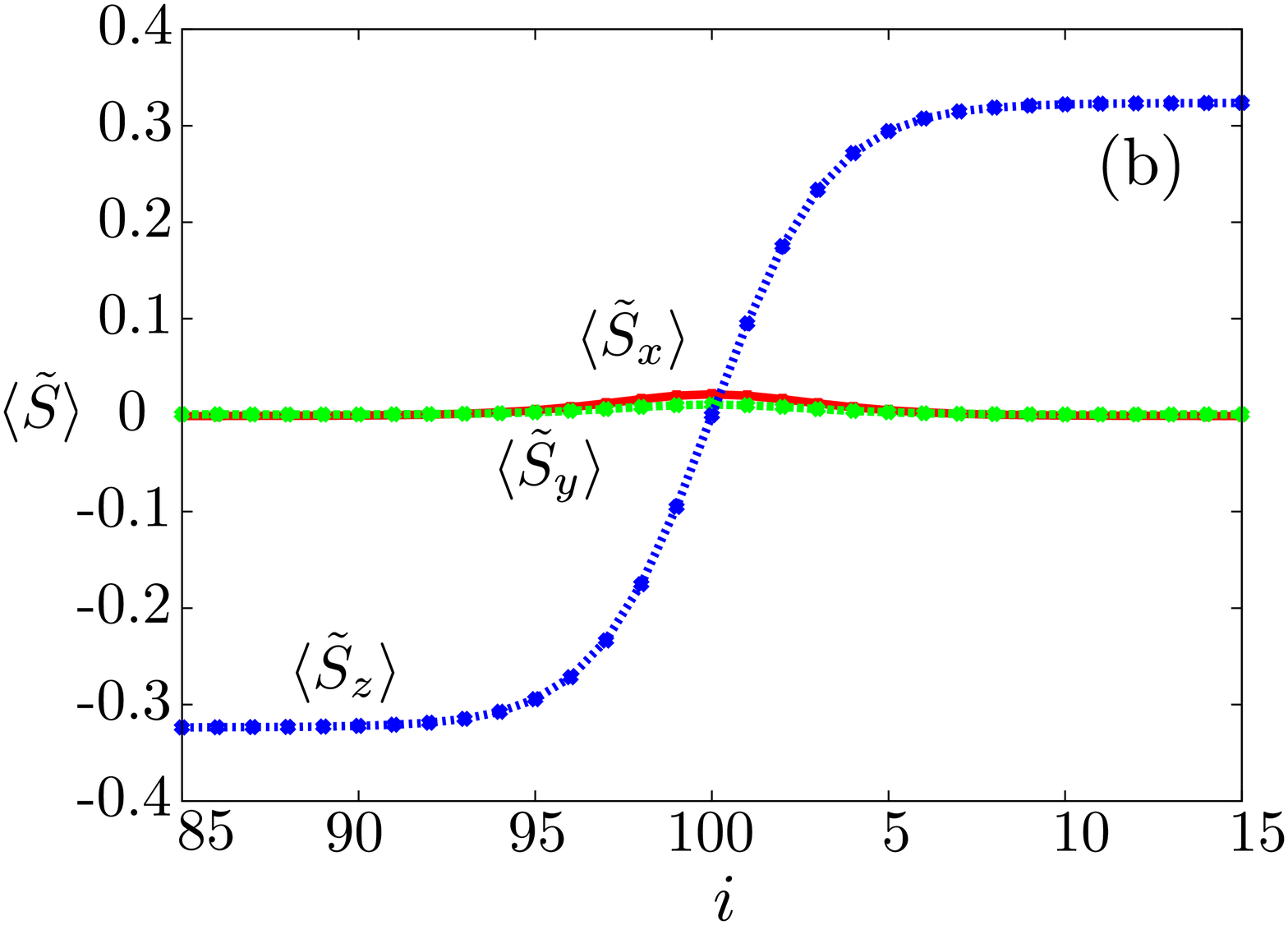}
\includegraphics[width=5.5cm,clip]{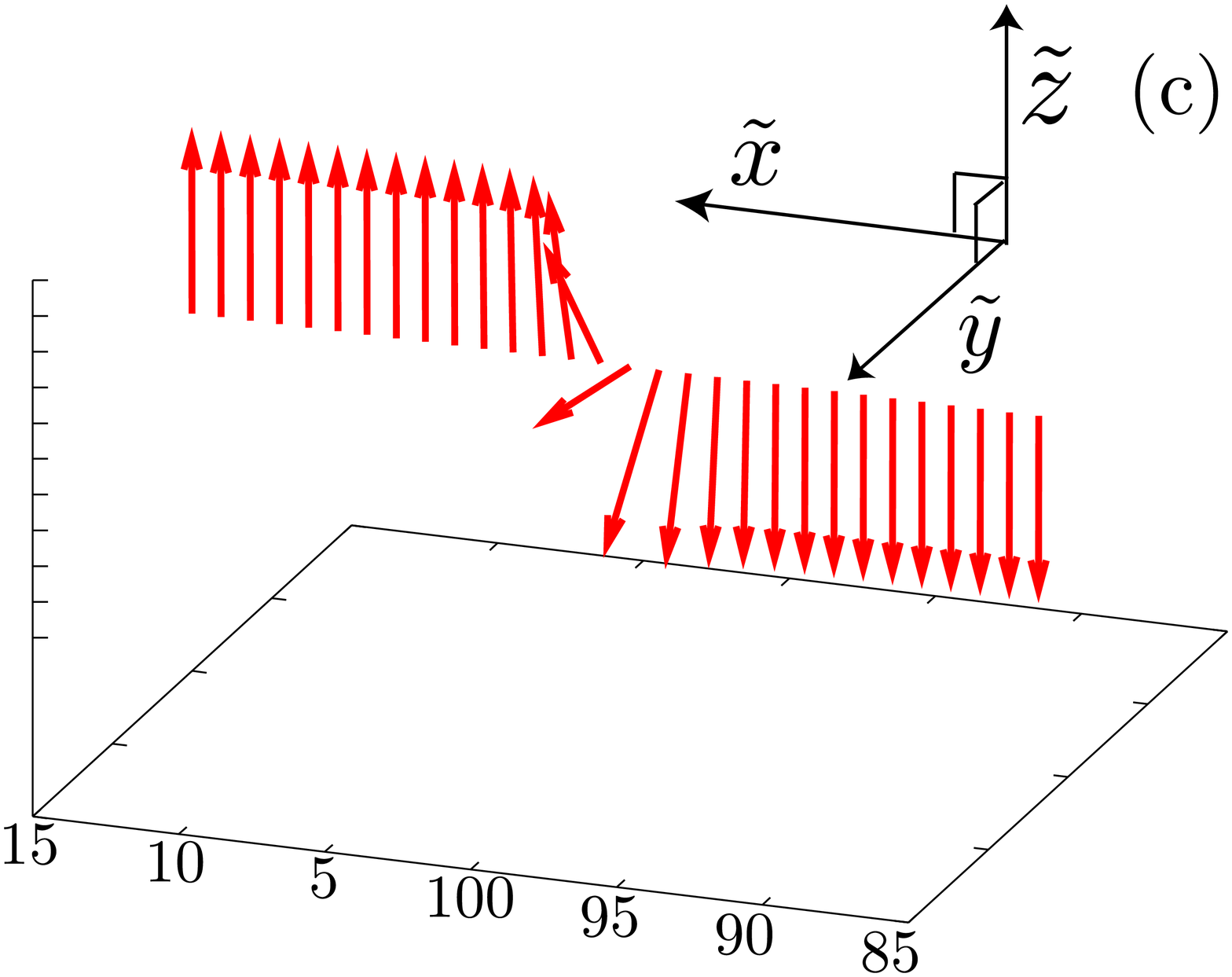}
\includegraphics[width=5.5cm,clip]{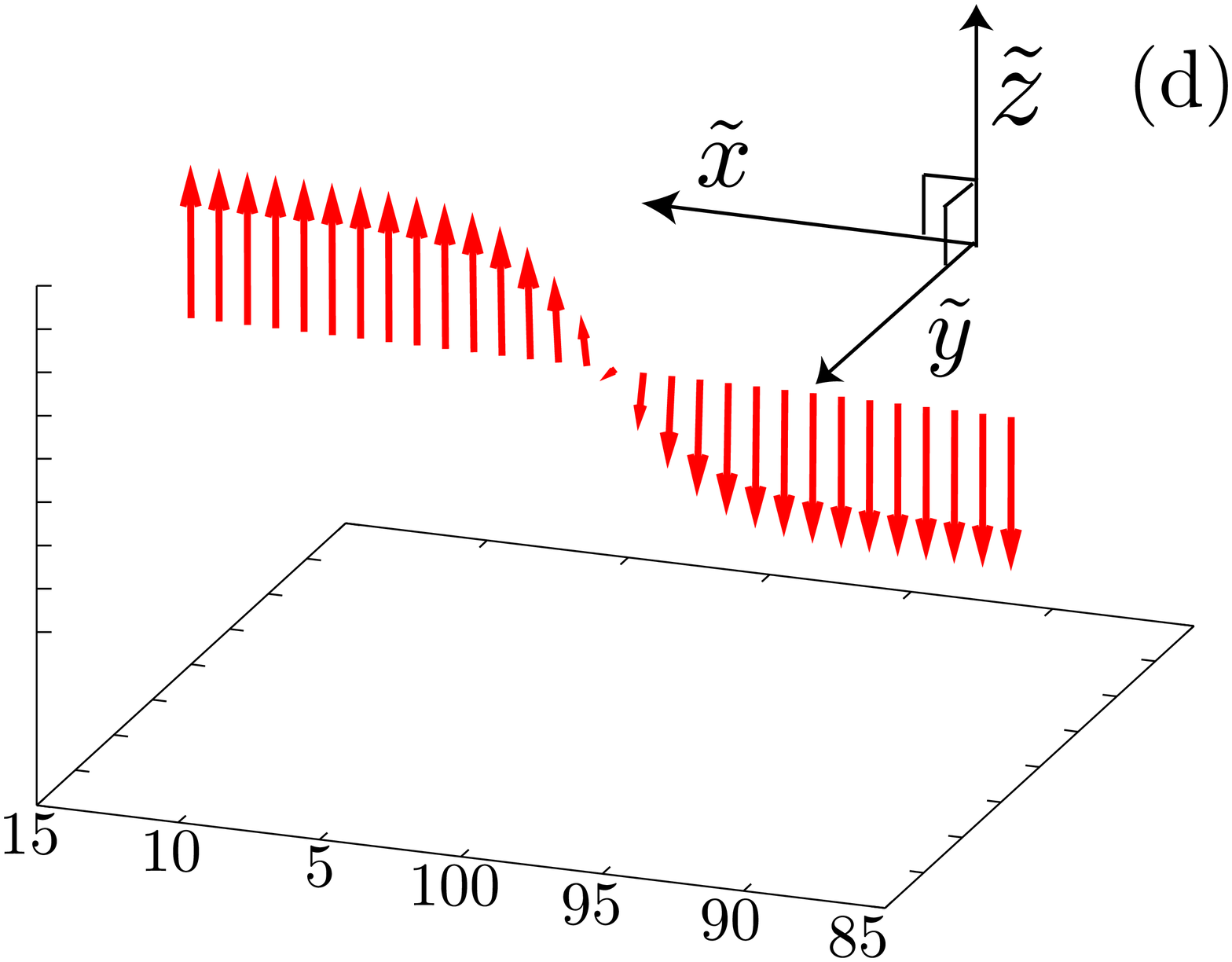}
\caption{
The spatial variations of $\langle {\tilde {\bf S}}_{i,j}\rangle$  
near the DW in the relative coordinate system
for $D=0.5$ and $\alpha=0.15$. $T=0.1T_C$ for (a) and (c), and 
$T=0.9T_C$ for (b) and (d).}
\end{center}
\end{figure}
As seen the region of type-B domain wall becomes larger as $T$ is increased 
or $D$ is decreased. 
This behavior can be understood as follows. 
Since we consider the quasi-one-dimensional system, the inter-chain 
exchange interaction $J\alpha$ is much smaller than the intra-chain
exchange $J$. 
The single-ion anisotropy energy $D$ was taken to be 
$J\alpha < D < J$, which seems to be a natural choice. 
Since the $D$ term favors the spin alignment normal to the strip, 
the loss of energy would become large for the broader width of the DW of type A. 
Then the width of the type-A DW is quite small irrespective of the 
temperature, actually a single site in most cases. 
On the contrary for the type-B DW, the width of the DW may become large
because $D < J$. 
As $D$ is further decreased or the temperature is increased, 
the width of DW can be larger and so the DW formation energy becomes 
smaller.\cite{chikazumi}
The spatial variations of $\langle {\tilde S}_{i,j}^\alpha \rangle$ 
(in the relative coordinate system) for $85 \leq i \leq 100$ and 
$1 \leq i \leq 15$ are depicted in Fig.6. 
The width of the type-B DW is actually smaller at lower $T$ for 
the same value of $D$. 
Here $\langle {\tilde S}_{i,j}^\alpha \rangle$ for $85 \leq i \leq 100$ are 
multiplied by -1, since the normal direction in the relative coordinate system 
is reversed between the sites with $i=100$ and $i=1$. 
We have also checked that the width becomes larger 
for smaller $D$ at the same temperature, as stated above. 
Thus the type-B DW becomes favorable compared to type-A DW, 
as $T$ is increased or $D$ is decreased. 
 
From Fig.5 we see that the phase transition from type-B DW to 
type-A DW occurs as the temperature $T$ is lowered. 
This results in an interesting behavior of the magnetization 
as a function of $T$. 
In Fig.7 the $T$ dependences of the magnetizations for 
$\alpha =0.15$ and $\alpha = 0.04$ are shown, where 
the total magnetization for each direction, $M_\mu$ ($\mu=x,y,z$), 
and its magnitude, $M_{tot}$, are defined as  
\begin{equation}\begin{array}{rl}
M_\mu = & \displaystyle \frac{1}{N_xN_y} \sum_{i=1}^{N_x} \sum_{j=1}^{N_y} 
\langle S^\mu_{i,j} \rangle , \\
M_{tot} = & \displaystyle \sqrt{M_x^2+M_y^2+M_z^2}.  
\end{array}\end{equation}
\begin{figure}[htb]
\begin{center}
\includegraphics[width=6cm,clip]{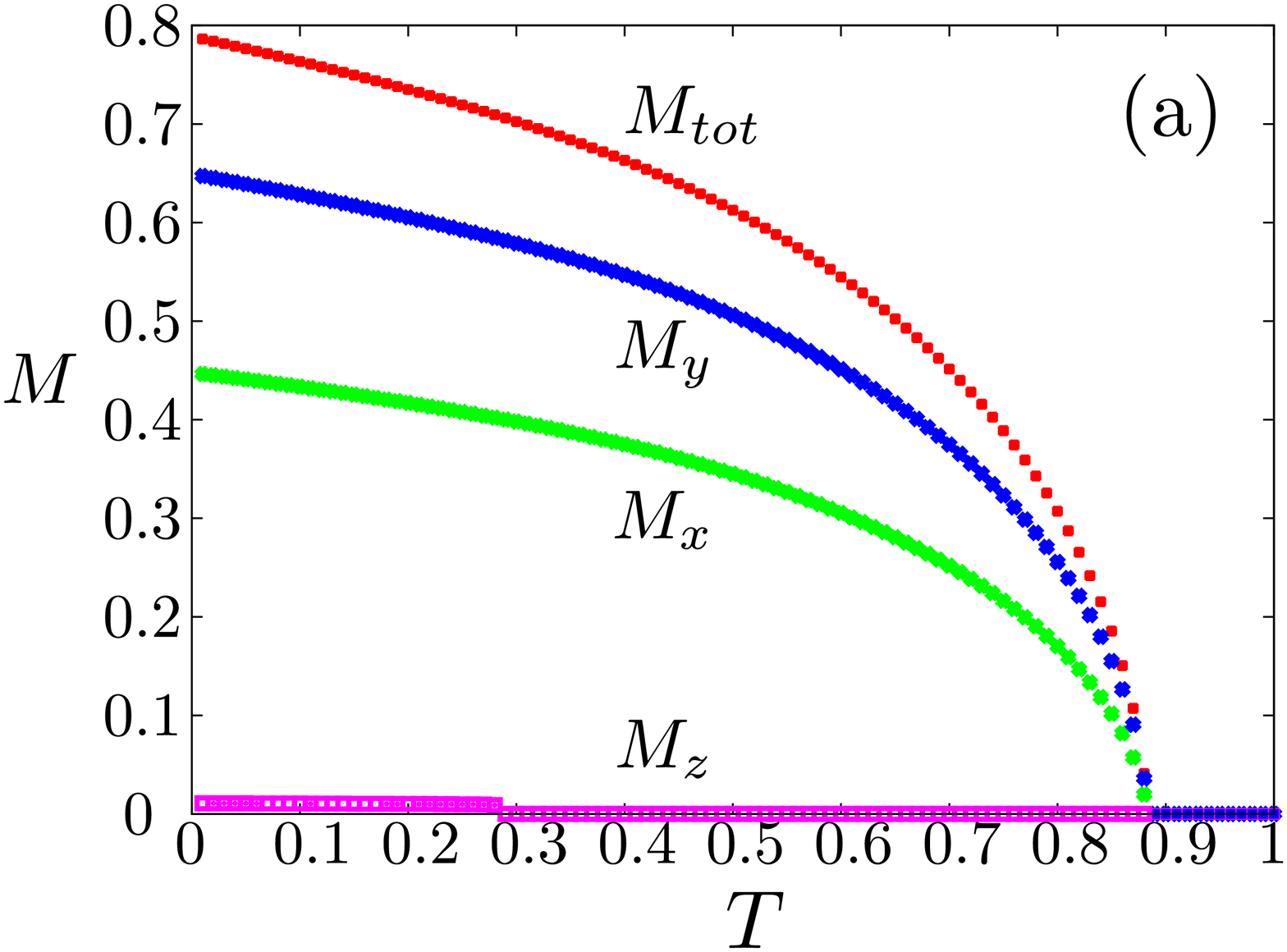}
\includegraphics[width=6cm,clip]{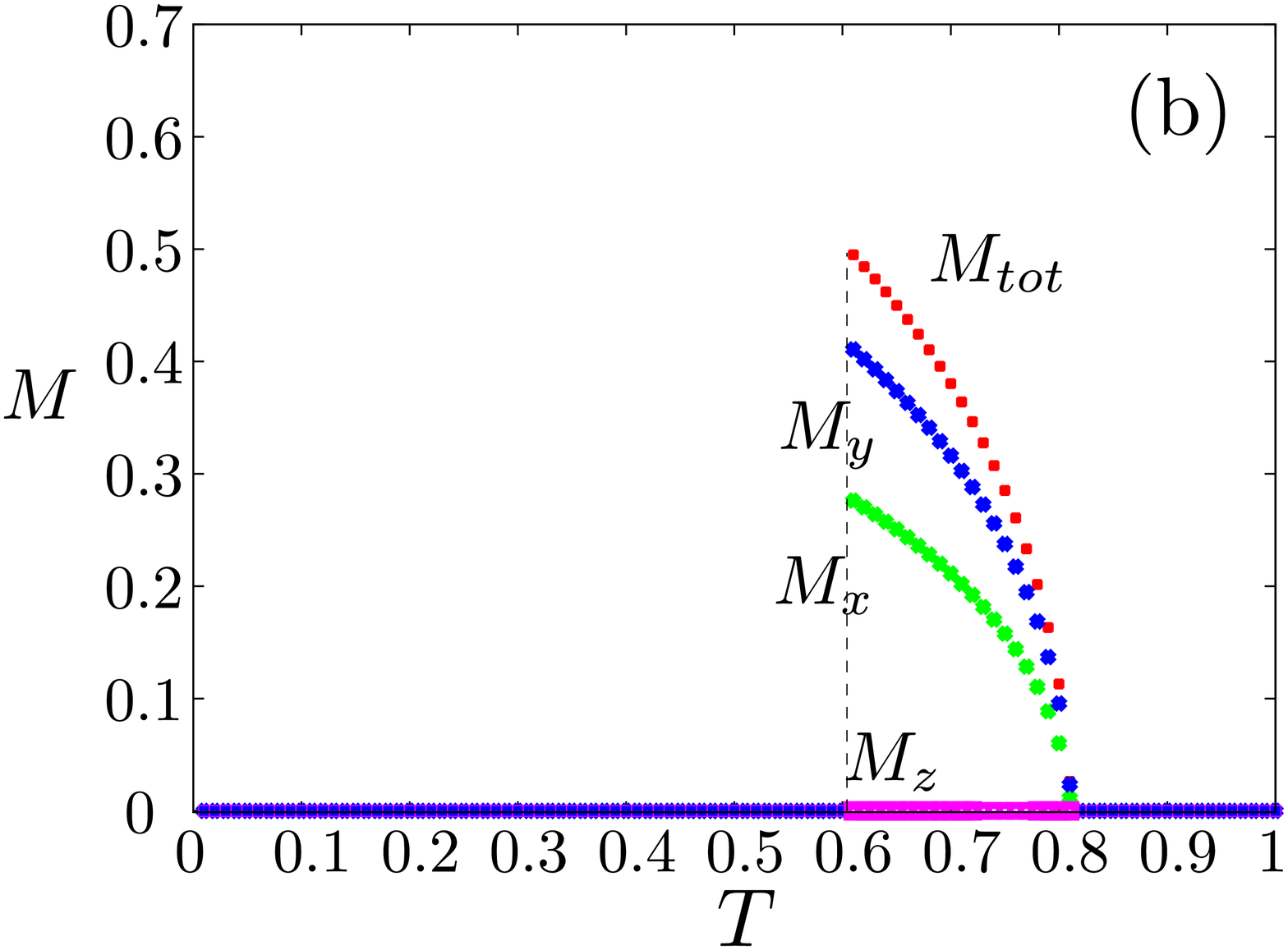}
\caption{
$T$ dependences of the magnetizations for $D=0.5$, 
(a) $\alpha =0.15$, and (b) $\alpha = 0.04$. 
The magnetization has a jump as indicated by the dashed vertical line. 
}
\label{dos}
\end{center}
\end{figure}
In the case of $\alpha=0.15$ the magnetization monotonously 
increases as $T$ is lowered. On the contrary the magnetization for $\alpha=0.04$ 
suddenly vanishes at $T \sim 0.6J$. This is because the 
phase transition from type-B DW to type-A DW occurs at this temperature.  
If the DW is of type A, the magnetizations in the $j$ th chain are exactly 
canceled by those in the $(N_y-j+1)$ th chain, though locally there 
are finite magnetizations. 
On the other hand, in the case of type-B DW there is no such 
cancellation and the total magnetization is finite, 
and it becomes larger as $T$ is lowered.  
In realistic systems this behavior of the magnetization 
would be observed with the hysteresis.
However, the detection of this behavior may be difficult because type-A and type-B 
DW states have globally different structures so that the pinning of DWs due to impurities 
or dislocations would disturb their motion more seriously than usual cases. 

The Curie temperature for the system with M\"obius geometry 
is only slightly lower than that for the planar system with the same parameters. 
For $D=0.5$, $\alpha=0.04$, $N_x=100$, and $N_y=10$, 
the former is $0.811J$ while the latter is $0.812J$. 
(The $T_C$ for the infinite plane with the same $D$ and $\alpha$ is $0.813J$.)
When $N_x$ and $N_y$ become larger the difference of $T_C$ will be smaller, 
since the local curvature of the M\"obius strip is decreasing. 
This implies that the Curie temperature for the M\"obius system is close to 
that for the flat system in general, at least in the mean-field approximation. 


\section{effect of external magnetic field}

Next we examine the effect of an external magnetic field ${\bf H}$ 
on the domain-wall structure. 
Here ${\bf H}$ is assumed  to be applied along the $y$ direction, 
{\it i.e.}, ${\bf H}$ threads the M\"obius strip. 
The $T$ dependences of the magnetizations are shown in Fig.8.  
Because of finite $H$, the magnetization is finite for $T > T_C(H=0)$ 
and gradually decreases as $T$ is increased. 
\begin{figure}[htb]
\begin{center}
\includegraphics[width=6cm,clip]{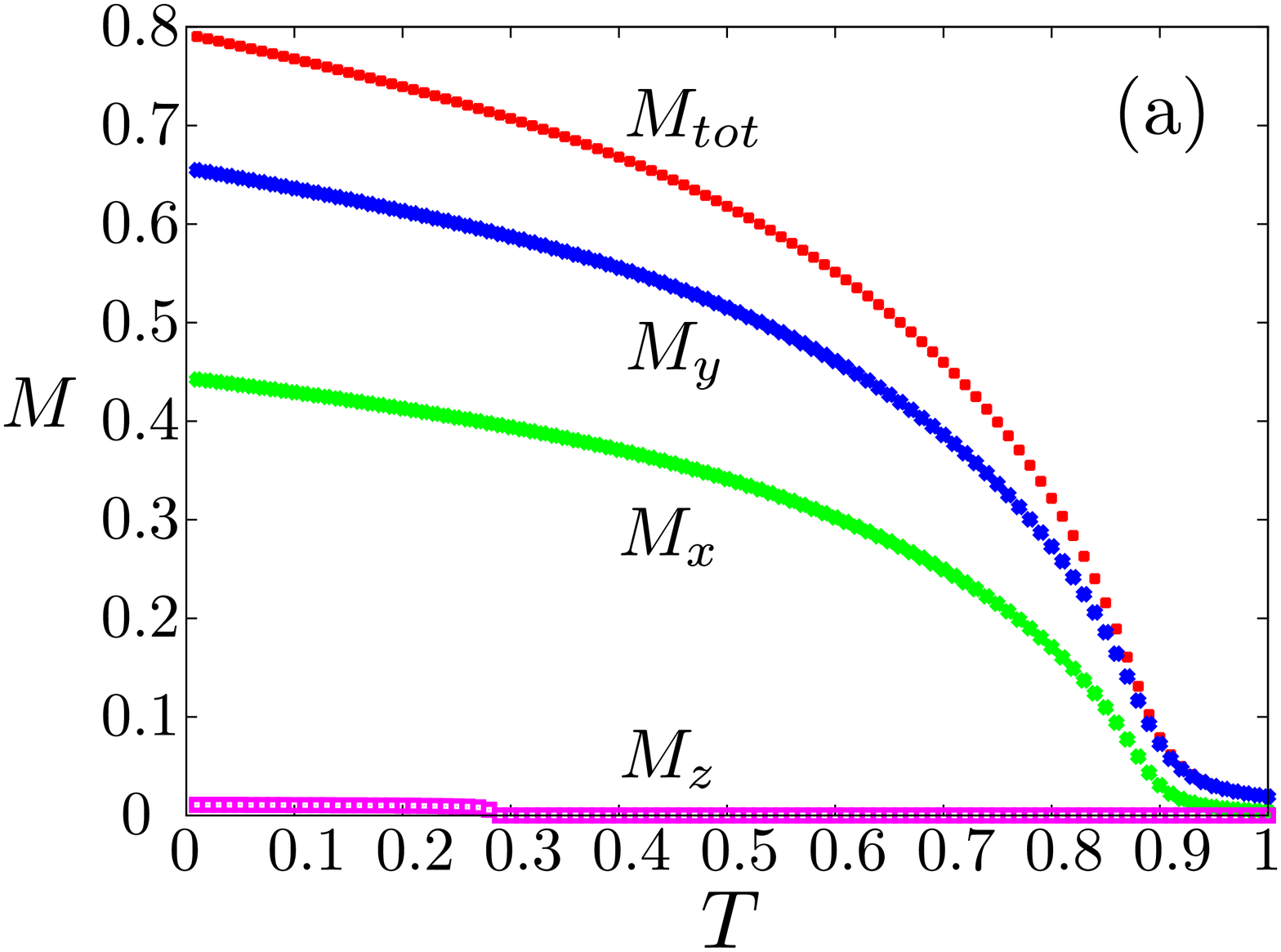}
\includegraphics[width=6cm,clip]{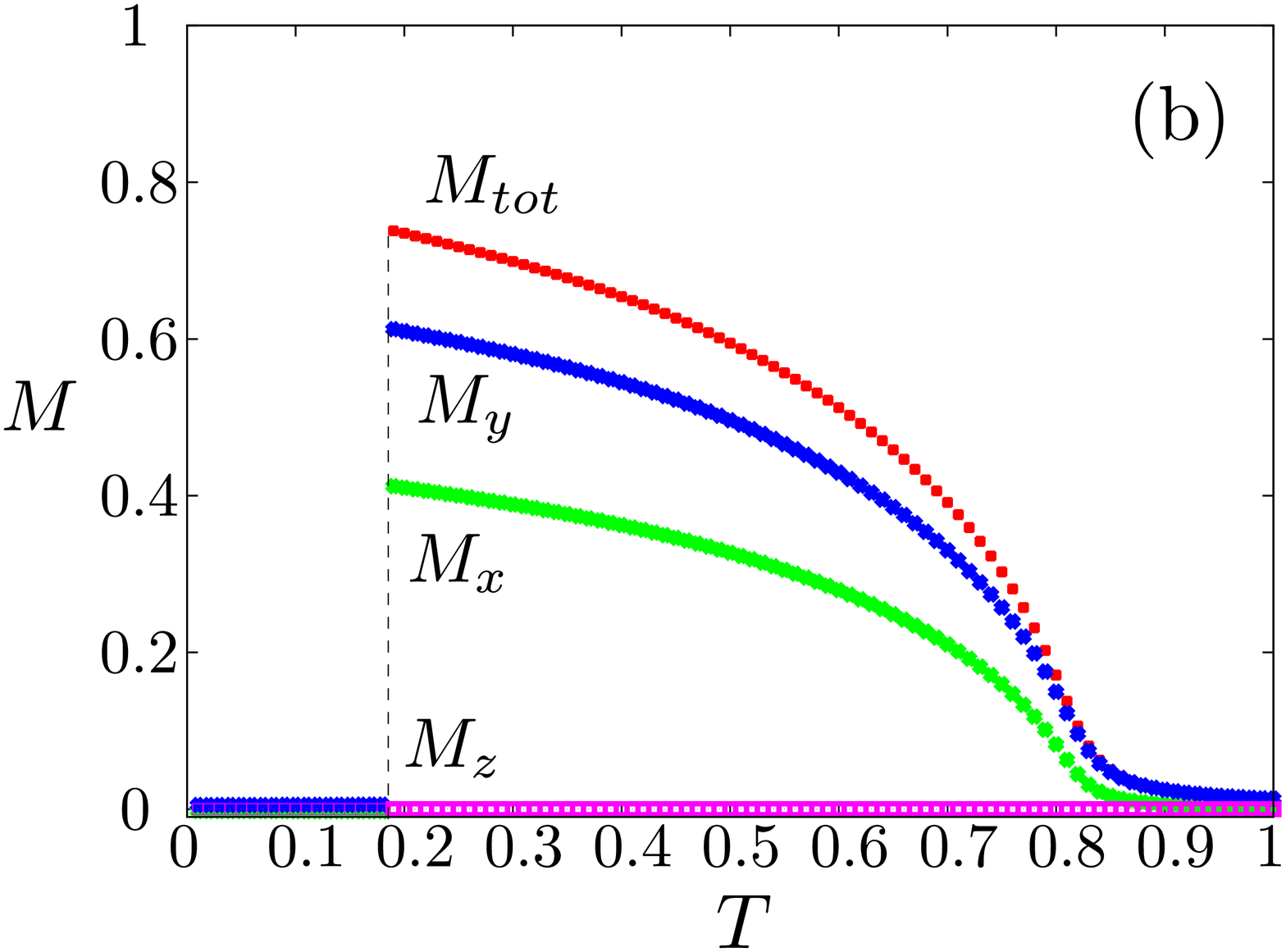}
\caption{
$T$ dependences of the magnetizations for $D=0.5$, $H=0.01$, 
(a) $\alpha = 0.15$, and (b) $\alpha=0.04$.
The magnetization has a jump as indicated by the dashed vertical line. 
}
\label{dos}
\end{center}
\end{figure}
\begin{figure}[htb]
\begin{center}
\includegraphics[width=6cm,clip]{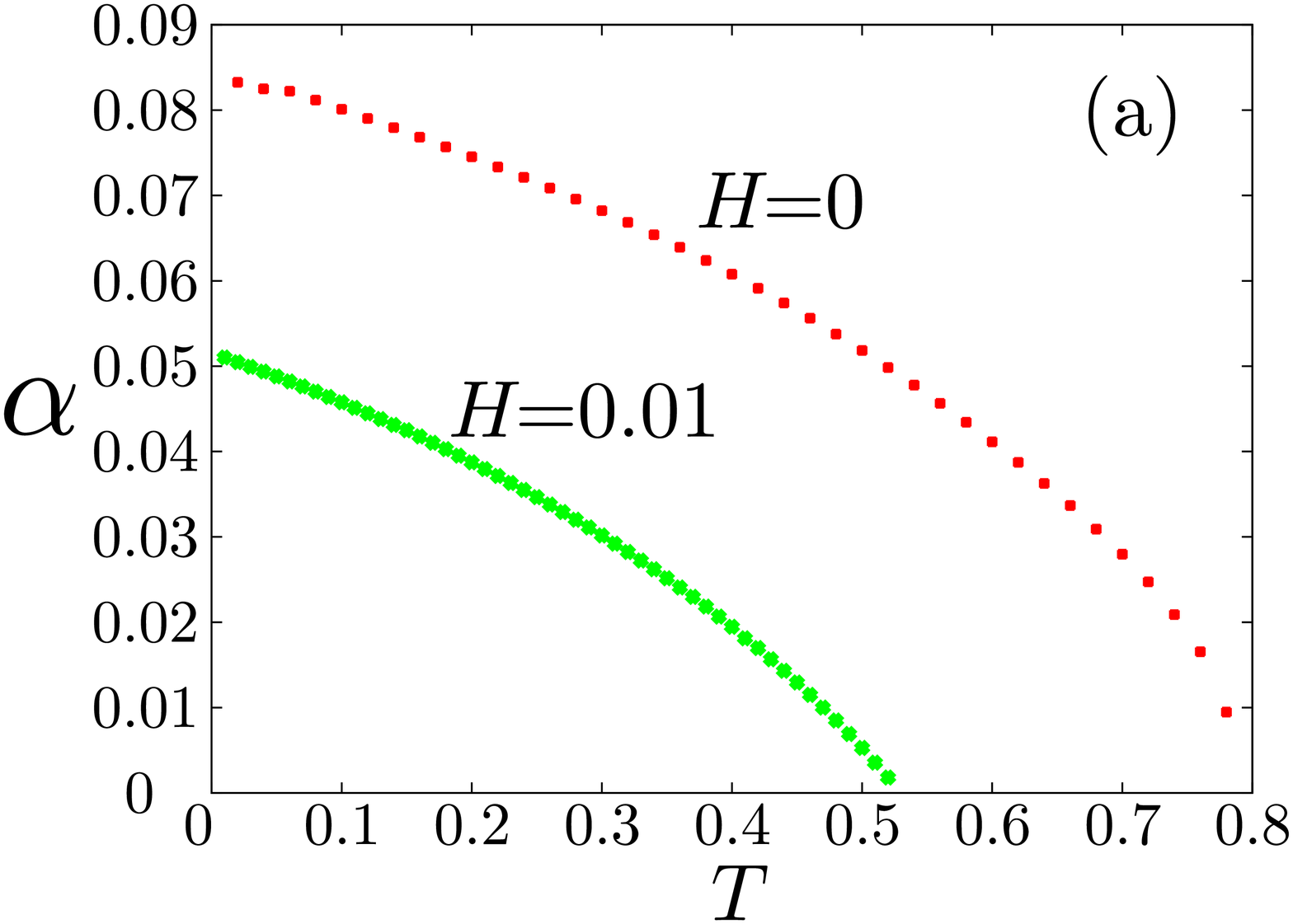}
\includegraphics[width=6cm,clip]{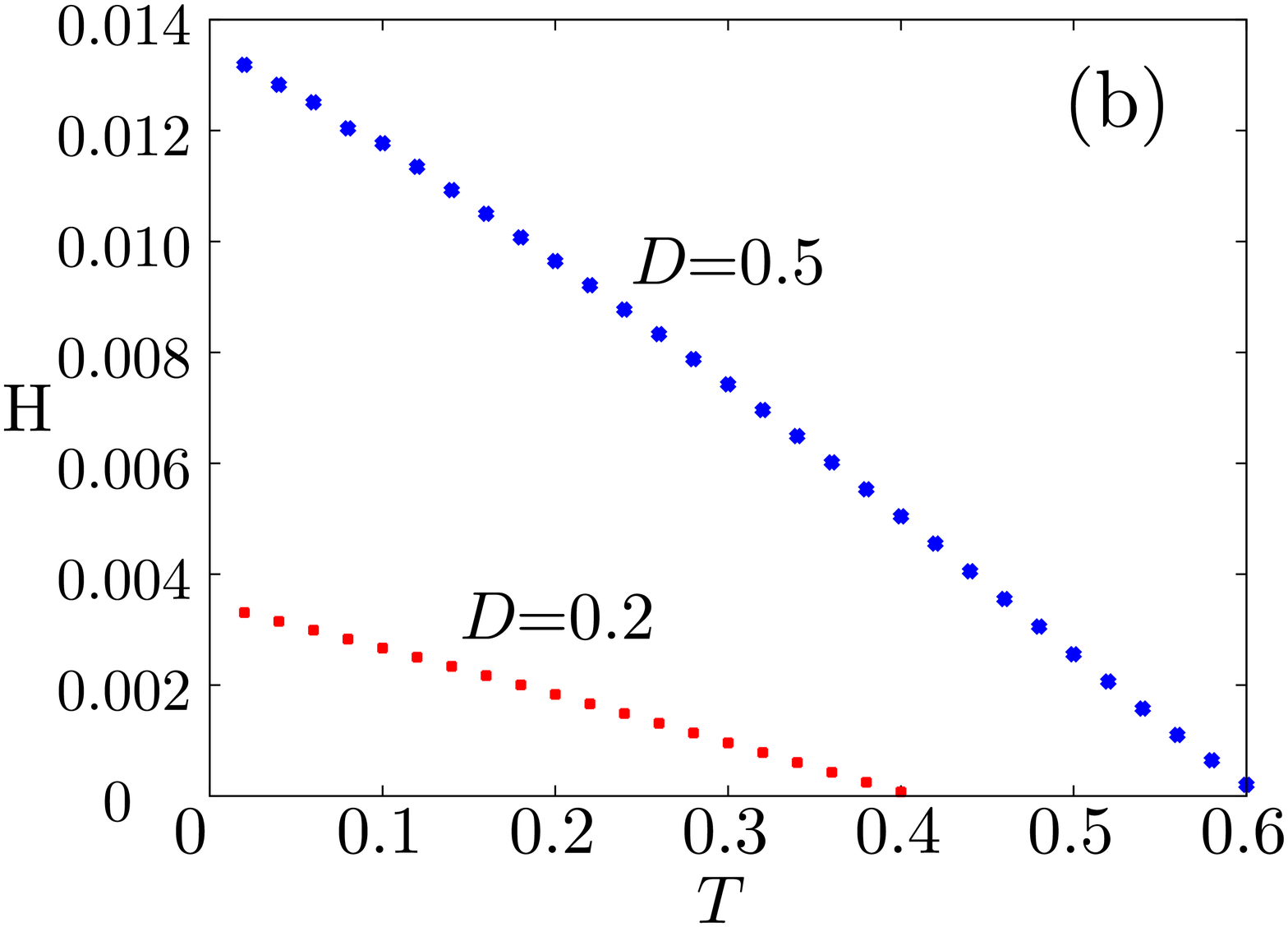}
\caption{
Phase diagram (a) in the plane of $\alpha$ and $T$ with $D=0.5$, and 
(b)  in the plane of $H$ and $T$ with $\alpha=0.04$. 
Curves are the boundaries between type-A and B domain walls, 
and the region close to the origin corresponds to the type-A domain wall. 
}
\end{center}
\end{figure}
\begin{figure}[htb]
\begin{center}
\includegraphics[width=6cm,clip]{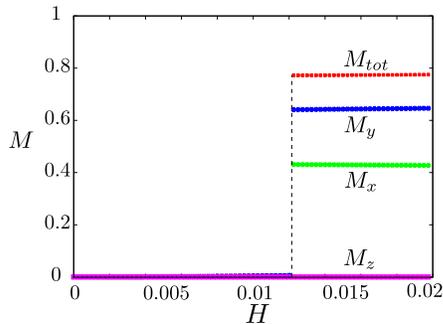}
\caption{
Magnetization as a function of the external field $H$ for $D=0.5$, 
$\alpha=0.04$, and $T=0.1T_C$.
The magnetization has a jump as indicated by the dashed vertical line. 
}
\end{center}
\end{figure}
For $\alpha=0.15$ [Fig.8(a)] the magnetization smoothly grows 
as $T$ is decreased for $T < T_C(H=0)$, while there is a discontinuity
in the case of $\alpha=0.04$ [Fig.8(b)]. 
This is again due to the transition from type-B to type-A DW, 
but the transition occurs at a much lower temperature compared to  
the case of $H=0$ [Fig.7(b)]. This is because the Zeeman coupling lowers 
the energy of the state with type-B DW due to its finite total magnetization 
and type-B state is more favored in the presence of the external field. 
For $H \not= 0$ the magnetization in the type-A state is finite 
but is very small. 
Next we show the phase diagram 
in the plane of $T$ and $\alpha$ for fixed value of $H$ [Fig.9(a)] 
and that in the plane of $T$ and $H$ for fixed $\alpha$ [Fig.9(b)]. 
In both Figs.9(a) and 9(b) the region of type-A DW shrinks when $H$ is applied. 
This can be understood similarly as the $T$ dependence of the magnetization. 
In the type-B state the magnetization is already finite without $H$, 
so that the energy can be gained due to the coupling to the applied magnetic field.  
Then the type-B state is more favorable compared to the type-A state 
in the presence of $H$.  
This resulted in a serious difference in  
response to the applied magnetic field for two types of DWs. 
The dependence of the magnetization on the external 
magnetic field is shown in Fig.10. 
There is again a jump of the magnetization as a function of $H$ 
because of the transition between different types of DWs. 
If we assume $J\sim$10meV, the transition point locates at around 
$H$=1T for the parameters used here. 

In the above we have assumed that ${\bf H}$ is applied along the 
$y$ direction. For the type-B DW, this direction is parallel to the domain wall. 
When ${\bf H}$ is applied along other directions, type-B DW will move 
to a different position at which the more Zeeman energy can be gained. 
This problem will be examined elsewhere.  

\section{summary} 

In summary, we have studied the structure of domain walls in 
ferromagnetic M\"obius strips. We found that there can be 
two kinds of domain walls whose relative stability is quite sensitive 
to the change in temperature and the applied magnetic field. 
The magnetizations will have interesting behaviors in that 
they have discontinuities as functions of both temperature 
and the magnetic field. 
Although ferromagnetic M\"obius strips have not yet been obtained, 
we expect that these systems will attract much interest both 
theoretical and experimental once they are synthesized.

\begin{acknowledgments}
M.H was financially supported by 
Grants-in-Aid for Scientific Research of Ministry
of Education, Science, and Culture. 
\end{acknowledgments}


\end{document}